# Machine learning for nocturnal diagnosis of chronic obstructive pulmonary disease using digital oximetry biomarkers

Jeremy Levy[1,2], Daniel Álvarez[3,4,5], Felix del Campo[3,4,5] and Joachim A. Behar[2]

[1] Faculty of Biomedical Engineering, Technion Institute of Technology, Haifa, Israel

[2] Faculty of Electrical Engineering, Technion Institute of Technology, Haifa, Israel

[3] Río Hortega University Hospital Valladolid, Valladolid, Spain

[4] Biomedical Engineering Group, University of Valladolid, Valladolid, Spain

[5] Centro de Investigación Biomédica en Red en Bioingeniería, Biomateriales y Nanomedicina (CIBER-BBN), Valladolid, Spain

**Abstract**

*Objective:* Chronic obstructive pulmonary disease (COPD) is a highly prevalent chronic condition. COPD is a major source of morbidity, mortality and healthcare costs. Spirometry is the gold standard test for a definitive diagnosis and severity grading of COPD. However, a large proportion of individuals with COPD are undiagnosed and untreated. Given the high prevalence of COPD and its clinical importance, it is critical to develop new algorithms to identify undiagnosed COPD, especially in specific groups at risk, such as those with sleep disorder breathing. To our knowledge, no research has looked at the feasibility of COPD diagnosis from the nocturnal oximetry time series.

*Approach:* We hypothesize that patients with COPD will exert certain patterns and/or dynamics of their overnight oximetry time series that are unique to this condition. We introduce a novel approach to nocturnal COPD diagnosis using 44 oximetry digital biomarkers and 5 demographic features and assess its performance in a population sample at risk of sleep-disordered breathing. A total of n=350 unique patients polysomnopgrahy (PSG) recordings. A random forest (RF) classifier is trained using these features and evaluated using the nested cross-validation procedure.

*Main results:* The RF classifier obtained F1=0.89±0.03 and AUROC=0.94±0.02 on the test sets. A total of 7 COPD individuals out of 70. No severe cases (GOLD 3-4) were misdiagnosed. Including additional, non-oximetry derived, PSG biomarkers only slightly improved the classifier performance from AUROC=0.94±0.02 to 0.95±0.01.

*Significance:* Our research makes a number of novel scientific contributions. First, we demonstrated for the first time, the feasibility of COPD diagnosis from nocturnal oximetry time series in a population sample at risk of sleep disordered breathing. We highlighted what digital oximetry biomarkers best reflect how COPD manifests overnight. The results motivate that overnight single channel oximetry is a valuable pathway for COPD diagnosis.

**1. Introduction**

Chronic obstructive pulmonary disease (COPD) is a highly prevalent chronic condition with a prevalence at 11.8% (95% confidence interval: 11.2–12.5) (Soriano *et al* 2020). COPD is characterized by persistent airflow limitation that is usually progressive and an enhanced chronic inflammatory response to noxious particles or gases in the airways and the lungs (Singh *et al* 2019). COPD is a major source of morbidity, mortality and healthcare costs. Etiological factors of COPD include aging, indoor and outdoor air pollution and history of smoking (Rice and Malhotra 2015). Suspicion of COPD is based on the clinical presentation of symptoms such as dyspnea, chronic cough or sputum production, reporting history of exposure to risk factors with mainly tobacco (Vogelheimer *et al* 2017). Diagnosis is confirmed if the ratio of forced expiratory volume within one second to forced vital capacity ($FEV_1/FVC$) is less than 0.70 in post-bronchodilator spirometry. Spirometry is the gold standard test for a definitive diagnosis and severity grading of COPD (Singh *et al* 2019).

*COPD an underdiagnosed condition*

A large proportion of individuals with COPD are undiagnosed and untreated (Diab *et al* 2018). Gershon et al. (Gershon *et al* 2018) reported 13.7% undiagnosed COPD cases in a Canadian adult (aged ≥ 40 years) random population-based sample (n=1,403 participants). This incidence was over 74,7% undiagnosed among the COPD patients in a Spanish adult population (n=9092) , as reported by (Soriano *et al* 2020). Given the high prevalence of COPD and its clinical importance, it is critical to develop new algorithms to identify undiagnosed COPD, especially in specific groups at risk, such as those with sleep disorder breathing.

*COPD and sleep-disordered breathing*

COPD is associated with other morbid conditions such as obstructive sleep apnea (OSA). For example, for patients with OSA, diagnosis of COPD is critical to identify the overlap syndrome (OVS) that consists of OSA and COPD concomitantly (Flenley 1985). OVS occurs in an estimated 1 in 10 patients having one of the two conditions (Malhotra et al 2018). The likeliness of developing additional serious conditions is greater than the likeliness with either disease alone (McNicholas 2017). Both OSA and COPD are highly prevalent diseases. Their coexistence leads to major social and healthcare-related consequences, particularly in the context of cardiovascular disease, as well as to an increased annual cost. Long-term clinical studies have found increased overall and cardiovascular mortality in OVS patients (Marin et al. 2008). Therefore, an early diagnosis of COPD is essential for effective treatment and a reduction in mortality of OSA patients. Unfortunately, in patients with suspected OSA, existing guidelines do not state the need for systematic respiratory functional assessment. A pulmonary evaluation would be particularly relevant for patients with smoking history, obesity, or those showing major respiratory symptoms, such as dyspnea (Lemarié *et al* 2010). In a recent perspective paper (Behar 2020) we motivated using overnight physiological recordings for the study, diagnosis and monitoring of non-sleep specific conditions. In the case of COPD, oximetry is of particular interest as it reflects respiratory function.

*Manifestation of COPD on nocturnal oxygen saturation time series*
Nocturnal desaturations are frequent in COPD patients, being more common in the most severe cases and particularly in patients with the chronic bronchitis phenotype. These drops in oximetry predominantly occur during rapid eye movement (REM) sleep and commonly show night-to-night variability (Buekers *et al* 2019). COPD and OSA are characterized by different hypoxemia models: OSA individuals show an intermittent pattern of desaturations during sleep. In advanced COPD patients, it is common to observe overnight chronic hypoxemia in individuals with no primary sleep disorders (Budhiraja *et al* 2015). It was reported that up to 70% of COPD patients with daytime saturations in the range of 90-95% had significant nocturnal hypoxemia (Chaouat *et al* 1997, Lewis *et al* 2009) and a lower mean overnight oxygen saturation as compared to controls (Valipour *et al* 2011). The overall prevalence of nocturnal desaturation in COPD patients was reported to vary from 27% to 49.2% (Fletcher *et al* 1987, Lewis *et al* 2009). Patients with OVS show more significant nocturnal desaturations (Lee and McNicholas 2011) than patients with COPD or OSA alone. (Chaouat *et al* 1995) noted greater nocturnal hypoxemia (lower mean $SpO_2$ in their study) in patients with OVS than patients with OSA alone. In (Sanders *et al* 2003), the odds ratio for desaturation below 85% for greater than 5% TST was approximately 20-fold greater in participants with OSA alone compared with those who had neither disorder. This number increased to about 30-fold in OVS patients. Thus, COPD may exert overnight disease specific oximetry patterns whether when found alone or concomitantly with OSA.

*The knowledge gap*

To our knowledge, no research has looked at whether it is possible to diagnose COPD from the nocturnal oximetry time series, either alone or concomitantly with another breathing disorder such as OSA.

*Hypothesis and objectives*

We hypothesize that patients with COPD will exert certain patterns and/or dynamics of their overnight oximetry time series that are unique to this condition. We introduce a novel approach to nocturnal COPD diagnosis using a machine learning (ML) model trained on oximetry digital biomarkers (Levy *et al* 2020) and assess its performance in a population sample at risk of sleep disordered breathing.

**2. Methods**

A block diagram describing the steps in elaborating the ML model is shown in Figure 1. The model takes the raw data as an input, performs a preprocessing step, extracts the digital oximetry biomarkers, and then performs the classification.

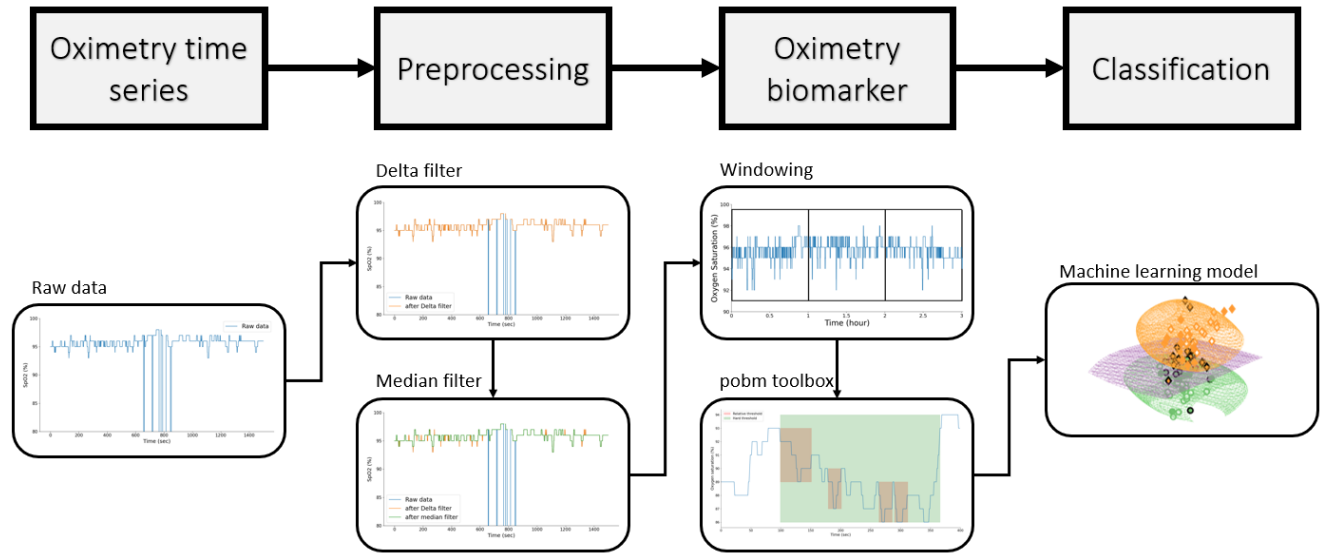

**Figure 1:** Block diagram describing the ML model elaboration.

*Database*

A total of 350 oximetry recordings were obtained during in-lab polysomnography (PSG). This database is described in the original work of (Andrés-Blanco *et al* 2017) which aimed at assessing the feasibility of automated OSA diagnosis from oximetry in patients with COPD. This database consists of 70 patients with confirmed COPD and 280 patients with no history of COPD in their medical records.. The latter group will be assumed to be "non-COPD". All participants showed high-to-moderate clinical suspicion of sleep disturbance breathing and they were referred for PSG in the sleep unit of the Rio Hortega University Hospital, in Valladolid, Spain. The COPD confirmed individuals were subjects aged 35 years and older, current or ex-smokers with a smoking history of at least 10 packs/years, referred to the Pneumology outpatient facilities due to symptoms indicative of the COPD disease. Complete pulmonary function assessment (Master screen PFT, Jaeger) was conducted for COPD patients, including pre-and post-bronchodilator spirometry, lung volumes, and lung diffusion capacity. The threshold used to confirm COPD from spirometry was $FEV_1/FVC < 0.70$. Standard in-lab PSG was carried out using a PSG E-series by Compumedics (Compumedics Limited, Victoria, Australia). Among patients with COPD (n=70), different subgroups were defined in terms of airflow limitations according to the global initiative for chronic obstructive lung diseases (GOLD) (Singh *et al* 2019): GOLD 1 (20.0%, n=14), GOLD 2 (65.7%, n=46), GOLD 3 (12.9%, n=9) and GOLD 4 (1.4%, n=1). GOLD 1 refers to a mild airflow limitation severity, while GOLD 4 means very severe airflow limitation (Vogelmeier *et al* 2017). Figure 2 presents a set of examples for each GOLD level and a non-COPD example. The $SpO_2$ times series are presented after preprocessing. For GOLD 1 and 3, the patients have also mild OSA. For GOLD 4, the patient has severe OSA. Figure 3*:* presents the repartition of the GOLD levels (1-4) and severity levels of OSA (mild, moderate and severe) (Thornton *et al* 2012) within the database. The median of the AHI among COPD patients in the database was 34.7, while among non-COPD patients it was 35.2.

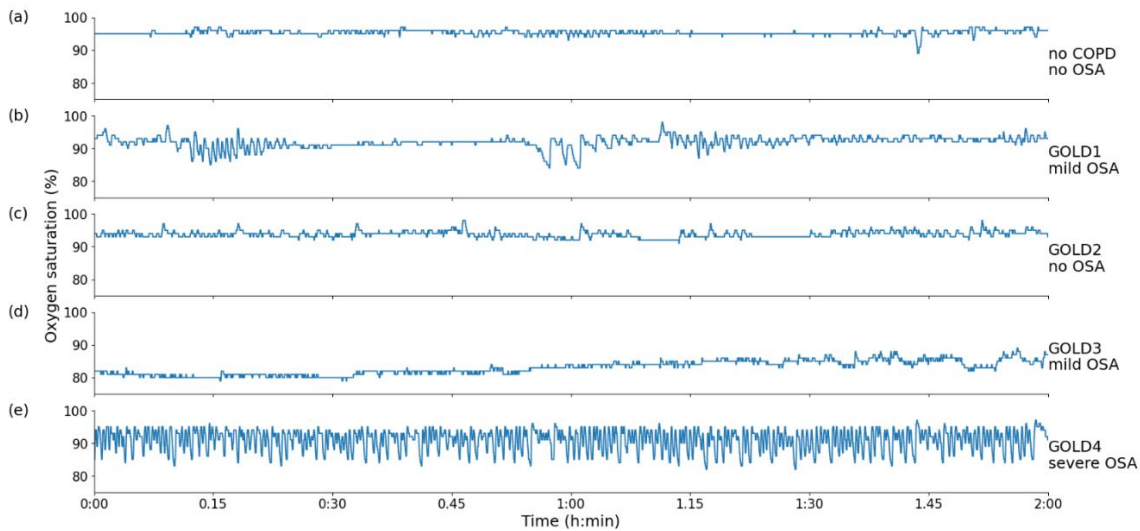

**Figure 2:** $SpO_2$ traces for all GOLD levels. On panel (a), $SpO_2$ time series of a patient diagnosed as non-COPD (healthy). On panel (b), patient with GOLD level 1 COPD. On panel (c), patient with GOLD level 2 COPD. On panel (d), patient with GOLD level 3 COPD. On panel (e), patient with GOLD level 4 COPD.

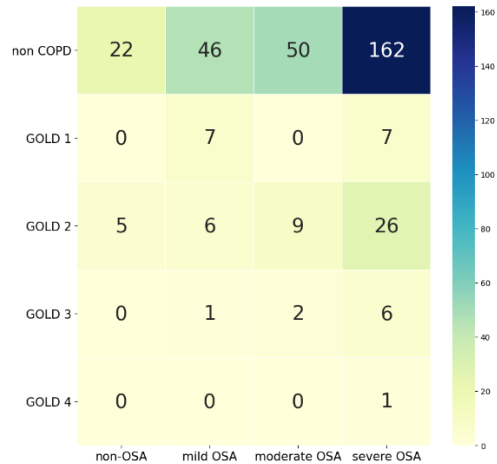

**Figure 3:** Patients in the study database as a function of their OSA and GOLD levels. Diagnosis and classifications are based on the guideline from (Thornton *et al* 2012, Ruoff and Rye 2016, Vogelmeier *et al* 2017).

*Preprocessing*

For the $SpO_2$ time series, the *Delta Filter* (Taha *et al* 1997, Levy *et al* 2020) was applied: all samples with values larger than 100 or smaller than 50 were considered non-physiological and excluded. Then a median filter of length 9 was applied to remove sharp changes (Deviaene *et al* 2019, Levy *et al* 2020). An example of preprocessing can be seen in Figure S1.

*Feature Engineering*

A total of 58 different features were computed (Table S1). These include 5 demographic features, 9 common, non-oximetry, PSG-derived features and 44 oximetry digital biomarkers engineered from the continuous $SpO_2$ time series (Levy *et al* 2020). The body mass index (BMI) was omitted, as it is redundant given weight and height are available as individual features. Table 1 presents the median and interquartile range for the demographic features. In addition, desaturation biomarkers were computed in two different ways namely, with a relative threshold and with a hard threshold. The relative threshold desaturation detector corresponds to the one used to compute the oxygen desaturation index (ODI) in sleep medicine. A hard threshold means that a desaturation is detected when the oximetry signal falls below a defined and constant threshold value – here taken as the median $SpO_2$. The intuition behind the hard threshold detector is that it may enable the model to detect the longer hypoxic events that are characteristic of COPD while the relative desaturation detector enables the identification of the shorter and more frequent desaturations observed in OSA patients. In the case of OVS, short desaturations may be embedded within those events, as can be seen in Figure 4 where there are four desaturations detected by the relative threshold (in red), whereas the hard threshold detected one longer desaturation (in green). In the case of a relative threshold, the maximum length of desaturation was set at 120 seconds. In the case of the hard threshold, there was no constraint on the desaturation length. Furthermore, all the oximetry biomarkers are computed over the full recording length and are added as additional "overall" features for individual window classification. The intuition is to give some context over the whole recording to improve the classification of individual windows. This process leads to a total number of 118 oximetry biomarkers, which combined with the demographic and PSG features results in 132 features overall.

*Statistical analysis*

To evaluate whether an individual feature was discriminative between the COPD and non-COPD groups, the Wilcoxon rank-sum test was used. Median and interquartile range are used for descriptive analysis of the features (Table 1). Violin plots are produced for the most discriminative features.

| Name | COPD ($n = 70$) | non-COPD ($n = 280$) |
|---|---|---|
| $Gender$ | Male : 61 (87%) | Male : 206 (74%) |
|  | Female: 9 (13%) | Female: 74 (26%) |
| $Age$ | 64.5 (59.3, 70.8) | 54.0 (45.0, 64.0) |
| $Weight$ | 84.0 (76.3, 94.0) | 82.0 () |
| $Height$ | 167.5 (163.0, 176.0) | 170.0 (163.0, 176.0) |
| $Smoking\ status$ | No smoker: 0 (0%) | No smoker: 83 (24%) |
|  | Smoker: 25 (36%) | Smoker: 139 (40%) |
|  | Ex-smoker: 45 (64%) | Ex-smoker: 58 (16%) |

**Table 1:** Median (MED) and interquartile range ($Quartile,\ 3^{rd}\ Quartile$) descriptive statistics of the population sample studied.

*Machine learning*

*Database split:* The database was separated into a training-validation set (80%) and test set (20%) using stratification with respect to the class COPD and non-COPD. Because of the imbalanced database, data augmentation was performed: each $SpO_2$ time series in the training set was decomposed into windows of two hours. For COPD patients, an overlap of one hour between consecutive windows was used. For non-COPD individuals, non-overlapping windows were used. The data augmentation procedure was used on the training set only. In the first step, individual windows of two hours were classified as COPD or non-COPD. As COPD is a chronic condition, a majority vote was then performed over the predicted labels of all the windows for a given recording in order to classify the patient as COPD or non-COPD.

*Models:* Four ML models were evaluated (Table 2): model 1 uses the demographic features only, model 2 uses the $SpO_2$ biomarkers extracted by the *pobm* toolbox developed in (Levy *et al* 2020), model 3 uses the $SpO_2$ biomarkers and the demographics features. Finally, model 4 uses all the features i.e. including other PSG features and is implemented in order to evaluate if there is value in using other standard PSG features versus oximetry alone.

*Feature selection:* Since model 1 has a low number of features, no feature selection step was applied. For models 2, 3 and 4 feature selection was performed using minimum redundancy and maximum relevance (mRMR) (Peng *et al* 2005). This algorithm aims to maximize the following operator:

$$\phi(S, c) = \frac{1}{|S|} \sum_{x_i \in S} I(x_i, c) - \frac{1}{|S|^2} \sum_{x_i, x_j \in S} I(x_i, x_j),$$

where $S$ is a subset of features, $I(x_i, c)$ is the information of the feature $x_i$ relative to the class $c$, $I(x_i, x_j)$ is the mutual information of features $x_i$ and $x_j$. This operator combines the Max-Relevance (first term), and the Min-Redundancy (second term). The set of features with the highest $\phi$ will be the set of features selected.

*Classifiers and cross-validation procedure:* For each model, two classifiers were trained; Logistic regression (LR) as a baseline model and Random Forests (RF) to evaluate the benefit of nonlinear classification. The Python library *scikit-learn* was used. Hyper-parameters were optimized using 5-fold cross-validation. A large random grid of hyper-parameters was searched (See Supplementary Note 2). For each iteration of the cross-fold, training examples were divided into train and validation set with stratification by patient. Because of the low number of patients in a single test set (20% of the overall database i.e. 70 patients), a nested cross-fold validation approach was taken. This means that 5-fold cross-validation was performed 5 times, each time on a different train-test split. This was done to report the median and variance performance of the models on the test sets.

|  | Demographics | PSG | All SpO$_2$ Biomarkers | Number of features | Selected features |
|---|---|---|---|---|---|
| **Model 1** | X |  |  | 5 | 5 |
| **Model 2** |  |  | X | 118 | 38 |
| **Model 3** | X |  | X | 123 | 35 |
| **Model 4** | X | X | X | 132 | 35 |

**Table 2:** Models being trained with respect to the input features used.

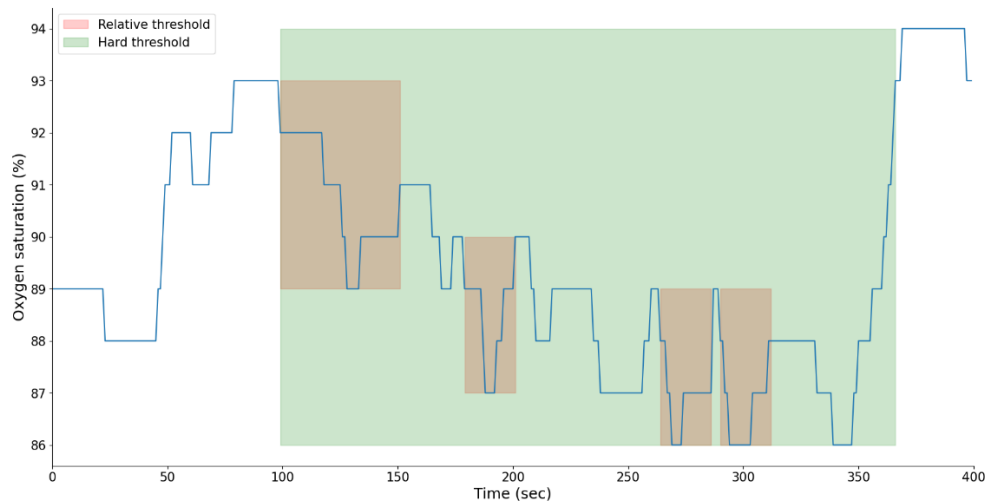

**Figure 4:** Hard and relative thresholds detector for desaturations.

## 3. Results

*Statistical analysis of the features*

The Wilcoxon rank-sum test rejected the null hypothesis for 115 out of 132 features. Tables S2, S3, S4, and S5 present the value of each set of features, with respect to the GOLD level. In particular, $MED$ ($p = 2.10 * 10^{-81}$) and $AV$ ($p = 8.67 * 10^{-75}$) yielded the lowest *p*-values. For 17 features the null hypothesis could not be rejected, e.g., height ($p = 0.365$) or $\Delta I$ ($p = 0.26$). The ranking of the 20 features with the lowest p-value can be seen in Figure S2. Additionally, a heatmap of correlated features is shown in Figure S3. This statistical analysis provides some insights about what features might be most discriminative between COPD and non-COPD patients. Figure 5 shows violon plots for the most discriminative features.

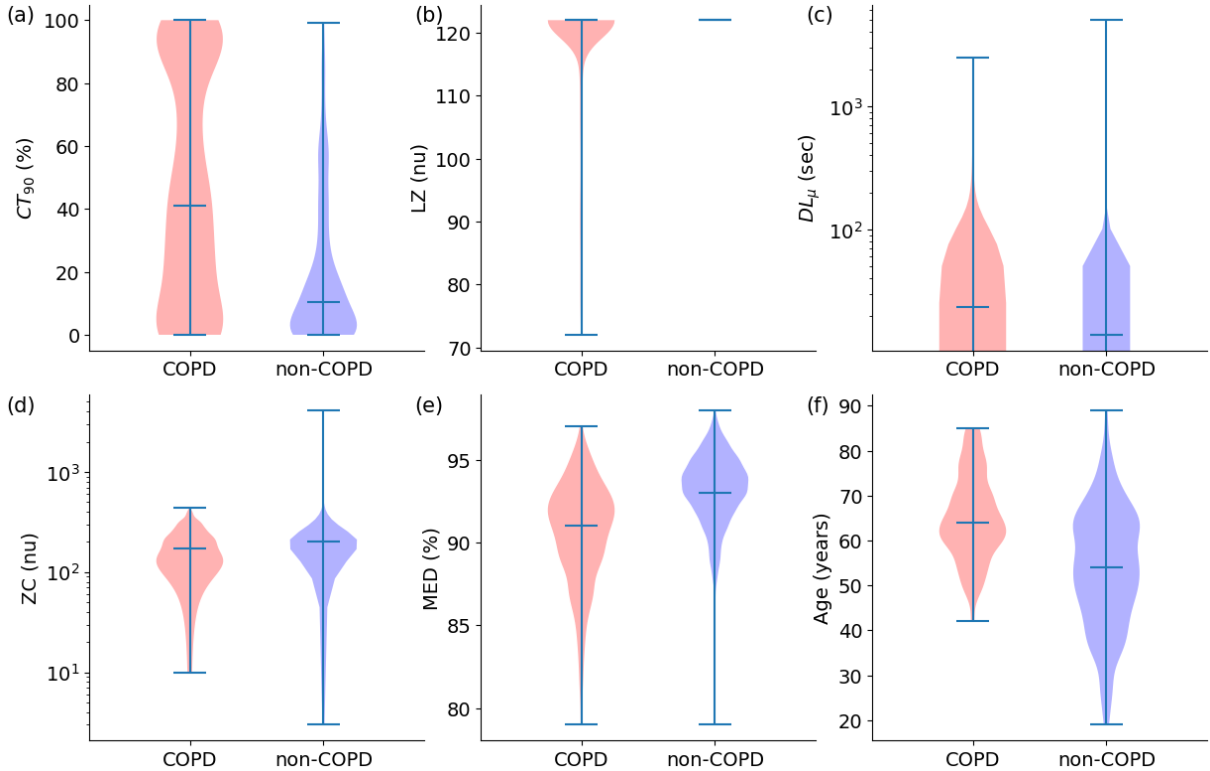

**Figure 5:** Violon plots for six discriminative ($p < 0.05$) features for the COPD and non-COPD groups. On panel (a): $CT_{90}$ (%); On panel (b): $LZ$ ($nu$); On panel (c): $DL_\mu$ ($sec$); On panel (d): $ZC$ ($nu$); On panel (e): $MED$ (%); On panel (f): $Age$ ($years$).

*Feature selection*

Given the limited number of examples (n=350) and the high number of features (up to 132 for model 4), it is important to reduce the dimensionality of the classification problem and see if this enables better performances to be reached. Using mRMR, a total of 38 features were selected for model 2 and 35 for models 3 and 4. The ranking of the selected features for models 2-4 are shown in Figure S4. The feature with the highest score is the most-relevant least-redundant feature.

*Classification*

The confusion matrix for the per window classification is provided in Table S6. The mean and standard deviation of the models' optimized hyperparameters are provided in Table S7. Table 4 presents the results on the test sets of the nested cross fold validation procedure, for models 1-4, for both RF and LR classifiers and for the per patient classification. Table 4 presents further performance measures for the RF classifier. The ROC curves are provided in Figure 6. Figure S5 presents the ROC curves for the classification per window. Table S8 presents the results on the test sets for the per window classification.

|  | RF | | | LR | | |
| --- | --- | --- | --- | --- | --- | --- |
|  | $AUROC$ | $F_1$ | $Kappa$ | $AUROC$ | $F_1$ | $Kappa$ |
| Model 1 | $0.69 \pm 0.13$ | $0.60 \pm 0.06$ | $0.39 \pm 0.12$ | $0.64 \pm 0.06$ | $0.56 \pm 0.10$ | $0.35 \pm 0.17$ |
| Model 2 | $0.80 \pm 0.09$ | $0.72 \pm 0.10$ | $0.62 \pm 0.05$ | $0.74 \pm 0.08$ | $0.61 \pm 0.13$ | $0.51 \pm 0.08$ |
| Model 3 | $0.94 \pm 0.02$ | $\underline{0.89 \pm 0.03}$ | $0.83 \pm 0.05$ | $0.82 \pm 0.13$ | $0.81 \pm 0.09$ | $0.64 \pm 0.06$ |
| Model 4 | $\underline{0.95 \pm 0.01}$ | $\underline{0.89 \pm 0.02}$ | $\underline{0.85 \pm 0.06}$ | $0.85 \pm 0.10$ | $0.81 \pm 0.10$ | $0.62 \pm 0.22$ |

**Table 3:** Per patients classification results for the outer-loop i.e. over the test sets for the RF and LR classifiers. The median and standard deviation of each performance measure over the five outer loops is presented.

|  | $Se\ GOLD1$ | $Se\ GOLD2$ | $Se\ GOLD3$ | $Se\ GOLD4$ | $Sp$ | $NPV$ | $PPV$ |
| --- | --- | --- | --- | --- | --- | --- | --- |
| Model 1 | $0.29 \pm 0.08$ | $0.64 \pm 0.06$ | $0.67 \pm 0.15$ | $\underline{1.00 \pm 0.00}$ | $0.85 \pm 0.02$ | $0.80 \pm 0.05$ | $0.42 \pm 0.15$ |
| Model 2 | $0.71 \pm 0.11$ | $0.76 \pm 0.06$ | $0.78 \pm 0.02$ | $\underline{1.00 \pm 0.00}$ | $0.79 \pm 0.06$ | $0.75 \pm 0.10$ | $0.70 \pm 0.12$ |
| Model 3 | $\underline{0.85 \pm 0.02}$ | $\underline{0.96 \pm 0.03}$ | $\underline{1.00 \pm 0.00}$ | $\underline{1.00 \pm 0.00}$ | $0.90 \pm 0.03$ | $\underline{0.96 \pm 0.02}$ | $0.83 \pm 0.08$ |
| Model 4 | $\underline{0.85 \pm 0.03}$ | $\underline{0.96 \pm 0.01}$ | $\underline{1.00 \pm 0.00}$ | $\underline{1.00 \pm 0.00}$ | $\underline{0.91 \pm 0.05}$ | $\underline{0.96 \pm 0.01}$ | $\underline{0.86 \pm 0.04}$ |

**Table 4:** Per patients classification results for the RF classifier and over the test sets. The median and standard deviation of each performance measure over the five outer loops are presented.

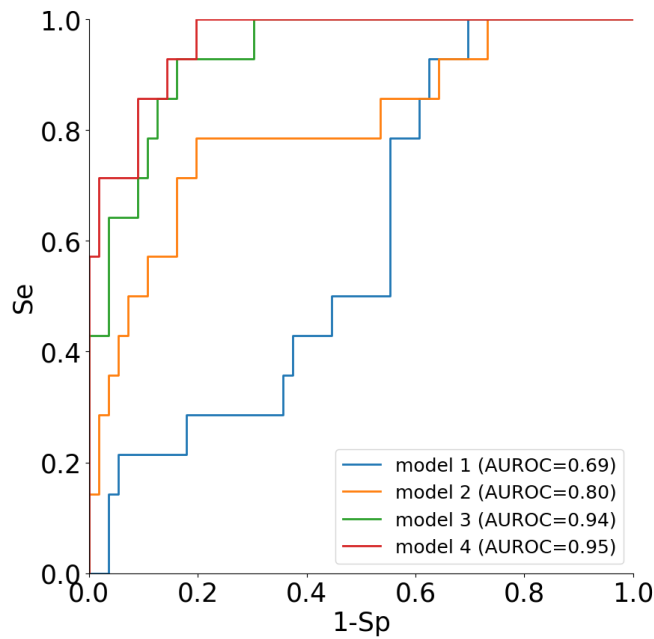

**Figure 6:** ROC curves for each of the 4 models, on the test set. Results are presented for the per patient classification.

## 4. Discussion

*Models performance*

Model 3 and 4 performed best with F1=0.89±0.03 and AUROC=0.94±0.02 and F1=0.89±0.02 and AUROC=0.95±0.01 respectively. The performance of model 3 was thus very close to model 4 which suggests that the diagnosis of COPD using single channel oximetry competes with a diagnosis that would use additional PSG biomarkers. In previous work, we have shown that single channel oximetry performed well in diagnosing OSA (Behar *et al* 2019, 2020). Combined with the present new results, the perspective is thus that using single channel oximetry it may be possible to diagnose both OSA and COPD remotely and thus provide the first single channel oximetry based diagnosis tool for both OSA and COPD and consequently OVS.

*Interpretability of features importance*

Figure 7: presents the feature importance ranking of the RF classifier for model 3 that is using oximetry biomarkers and demographics. The two most important features were demographic features, namely Age and Smoking Status. It is known that smoking is a risk factor that is highly important in COPD. For age, we observed that our COPD population was older than the OSA population (Table 5). The LZ complexity measure is ranked third. Figure 5 highlights that the COPD patients had the LZ biomarker with median and interquartile range (Q1-Q3) of 122 (104-122), whereas for non-COPD it was 122 (122-122). In the case of non-COPD patients with OSA, the repeated short desaturations represent a high degree of variations in the time series which is reflected by a high LZ feature value. The fourth most important feature is $CT_{90}$, the cumulative time under the 90% baseline. This feature captures the long hypoxic events in the signal. Previous research such as the one of (Lewis *et al* 2009) had also reported a high $CT_{90}$ in COPD patients. In addition, a number of desaturations features ranked high (6 in the top-15 features), both when using the relative ($DS_\sigma, DA100_\mu$) and hard ($DA100_\sigma, DDmax_\sigma$) desaturation thresholds. This reflects that the model relies on the desaturations slope and area in order to make the prediction. Figure 8 shows an example of desaturations characteristic of a COPD patient and a non-COPD patient with OSA. For the latest, the slope and area of the desaturations are close from each other. Indeed, many short consecutive desaturations can be observed. In this case, the features $DS_\sigma, DA100_\sigma, DDmax_\sigma$ will have low values. For the COPD patient, a single desaturation with greater area can be seen. In the case of COPD, the desaturations have a more variable length which will increase the standard deviation of the slope and the area of the desaturations which will lead to higher values of $DS_\sigma, DA100_\sigma, DDmax_\sigma$. Finally, overall features ($AC, MED, AV$) had a high feature importance which reflects that the classifier harnesses contextual information from the overall recording. Figure S6 presents the ranking of feature importance for the 4 models.

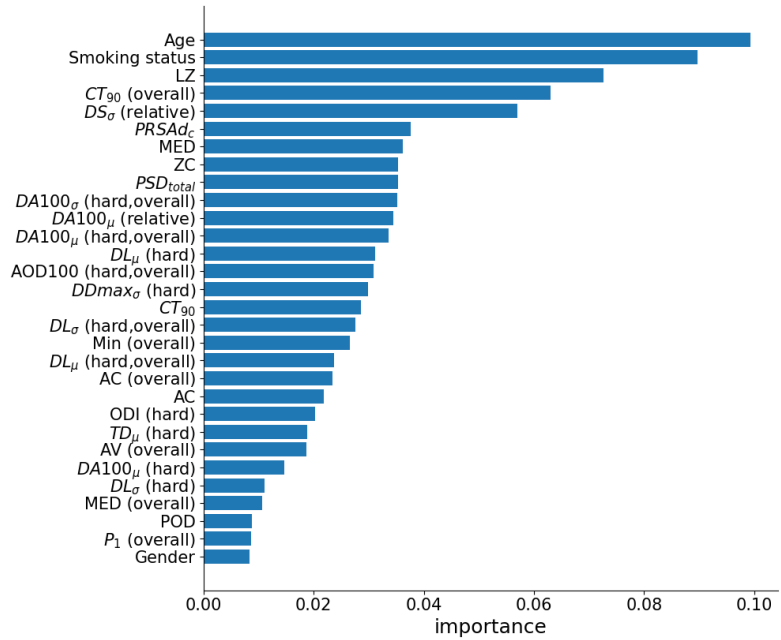

**Figure 7:** Feature importance of model 3, determined by RF classifier, for the 30 features with the highest score.

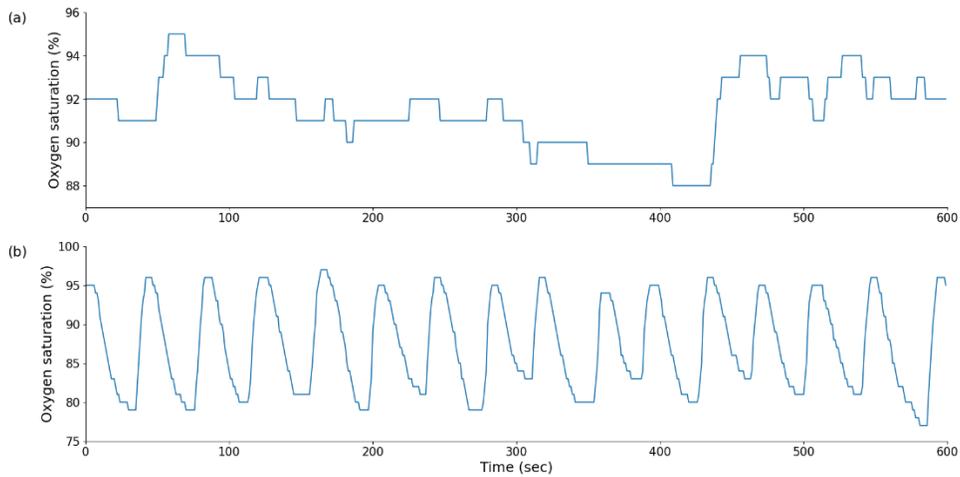

**Figure 8:** Example of desaturation characteristics. On panel (a), a COPD patient. On panel (b), a non-COPD patient with OSA.

*Error analysis*

Figure 9 presents the confusion matrix over the test sets, per patient. Most of the false positives (FP) have severe OSA (12 out of 18) which highlights that the classifier may be confused by the effect of repetitive desaturations that may "look like" longer desaturations that are characteristic of COPD. Secondly, all the false negatives (FN) belong to GOLD level 1and 2. No severe COPD cases (GOLD 3-4) were missed by the classifier. We noted that all individuals in the COPD group without OSA were correctly classified (5/5). The per window mean of the most important features for the TN, TP, FP and FN are summarized in Table 5. The model missed COPD especially for young patients (57.4 years against 65.4 years) and with a low $CT_{90}\ (overall)$ (42.0% against 46.8%) reflecting a lower number of hypoxic events. The FP were relatively older than the non-COPD patients in the database. Thus age may be misleading the classification of some examples.

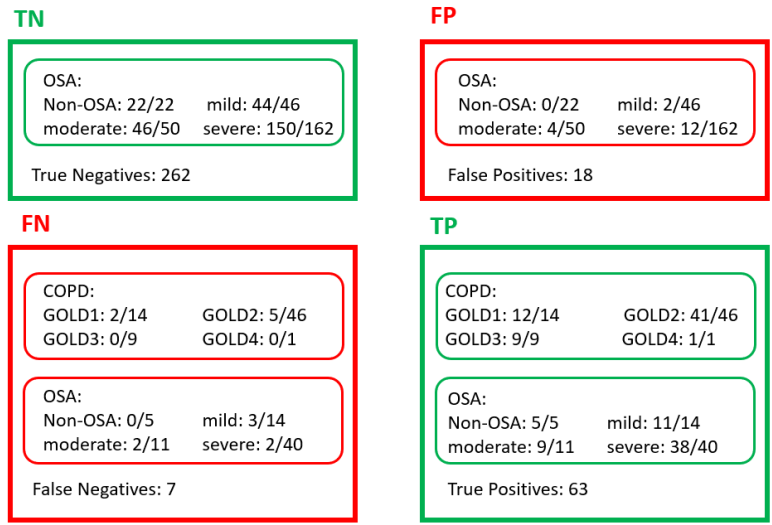

**Figure 9:** Confusion matrix for the test sets and model 3. The OSA and GOLD levels are specified.

| Mean | TP | FN | TN | FP |
|---|---|---|---|---|
| $Age$ | 65.4 | 57.4 | 53.1 | 59.6 |
| $Smoking\ status$ | 2.0 | 1.5 | 1.0 | 1.2 |
| $CT_{90}\ (overall)$ | 46.8 | 42.0 | 14.2 | 18 |
| $LZ$ | 118 | 121 | 122 | 122 |
| $DS_\sigma(relative)$ | 0.08 | 0.06 | 0.07 | 0.09 |

**Table 5:** Mean of most important features, for all the COPD database, COPD misdiagnosed windows, all the non-COPD database and then non-COPD misdiagnosed.

*Limitations*

The non-COPD group might actually contain some individuals with COPD, although there was no previous history of COPD in these patients medical record, i.e., neither symptoms nor exposure to risk factors, which are needed to suspect COPD and refer for spirometry according to the guideline (Vogelmeier *et al* 2017). Yet, this represents the main limitation of our work and motivates furthering this research by recording a new cohort where all the population sample undergoes a spirometry test.

**5. Conclusion**

Our research makes a number of novel scientific contributions. First we demonstrated, for the first time, the feasibility of COPD diagnosis from nocturnal oximetry time series in a population sample at risk of sleep-disordered-breathing. We highlighted what digital oximetry biomarkers best reflect how COPD manifests overnight. In particular $CT_{90}, LZ$ and $DS_\sigma$ were found to be the most discriminative. Finally, we show that including additional PSG biomarkers only slightly improves the classifier performance. This motivates single channel oximetry is a valuable option for COPD diagnosis.

**Acknowledgments:** We acknowledge the financial support of the Technion Machine Learning & Intelligent Systems Center. This work has been partially supported by 'CIBER en Bioingeniería, Biomateriales y Nanomedicina (CIBER-BBN)' through 'Instituto de Salud Carlos III' co-funded with FEDER funds.
**Competing interests:** JB holds shares in SmartCare Analytics Ltd. The remaining authors declare no competing interests.

# Supplementary material

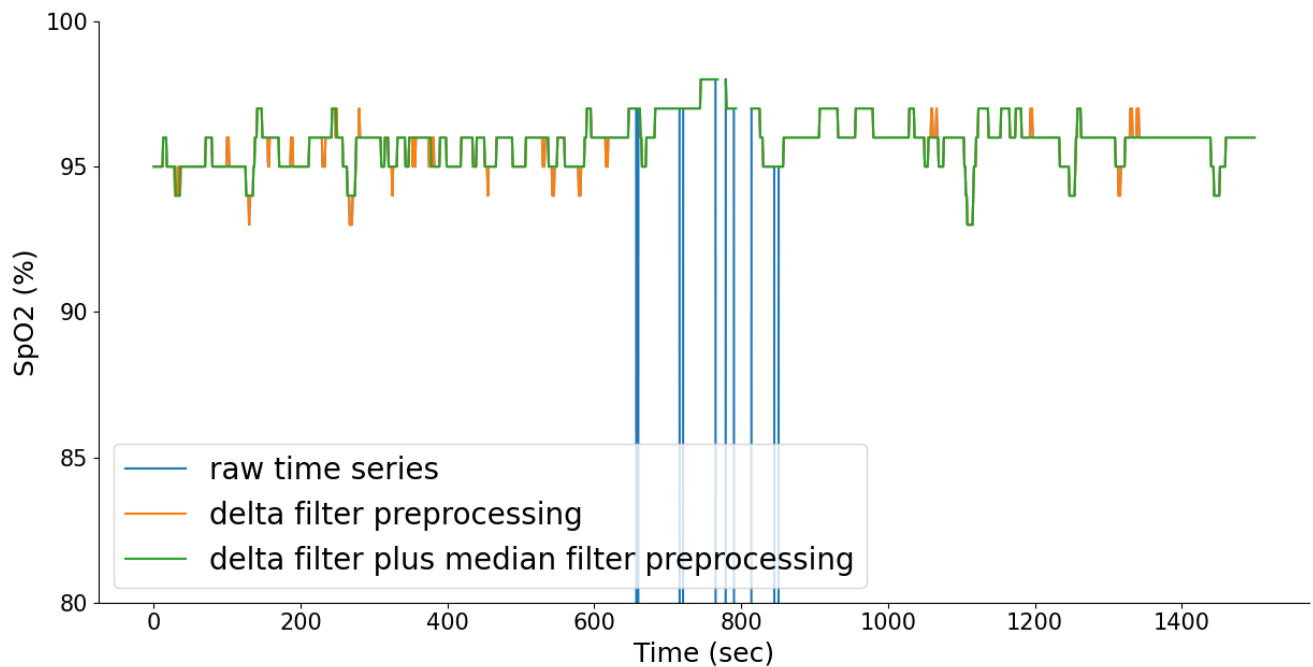

**Figure S1:** Example of the preprocessing step of the oximetry time series.

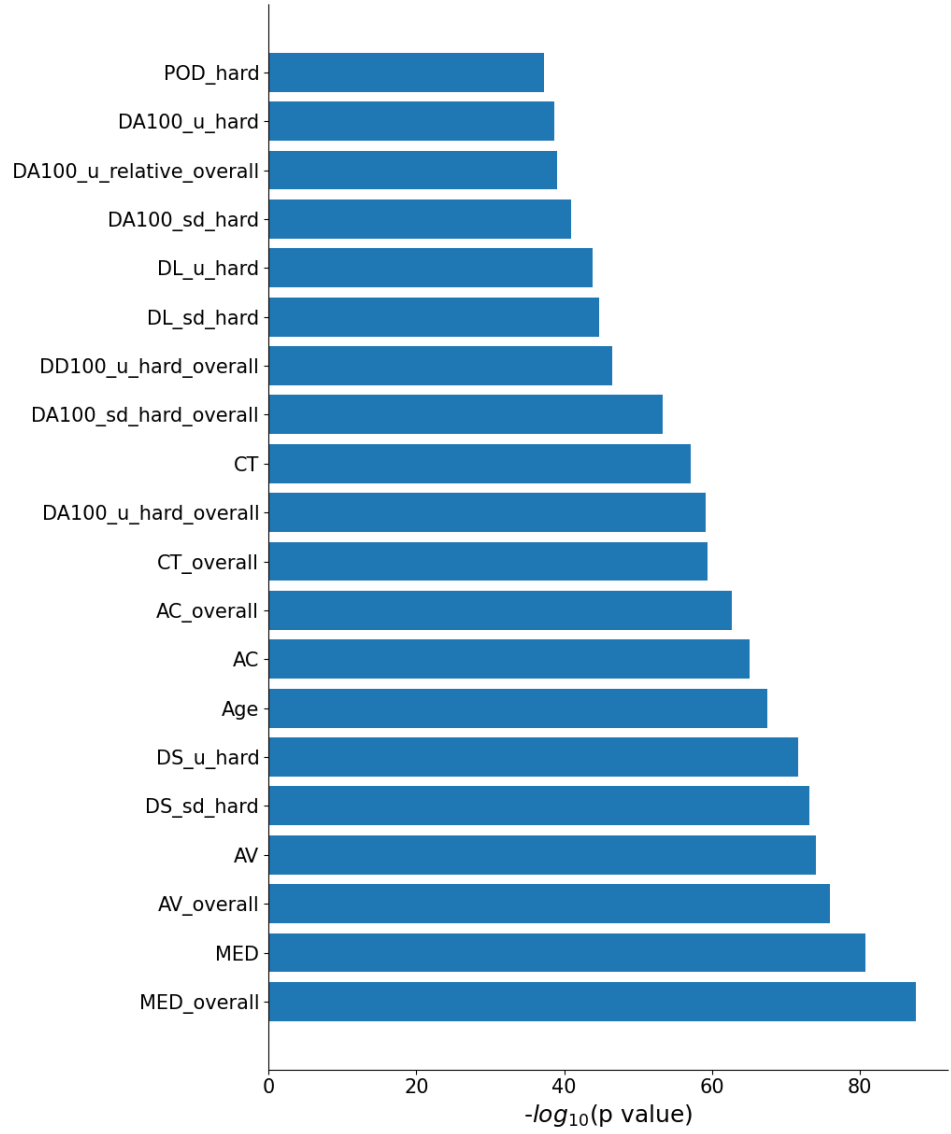

**Figure S2:** Statistical analysis: the 20 best p-values.

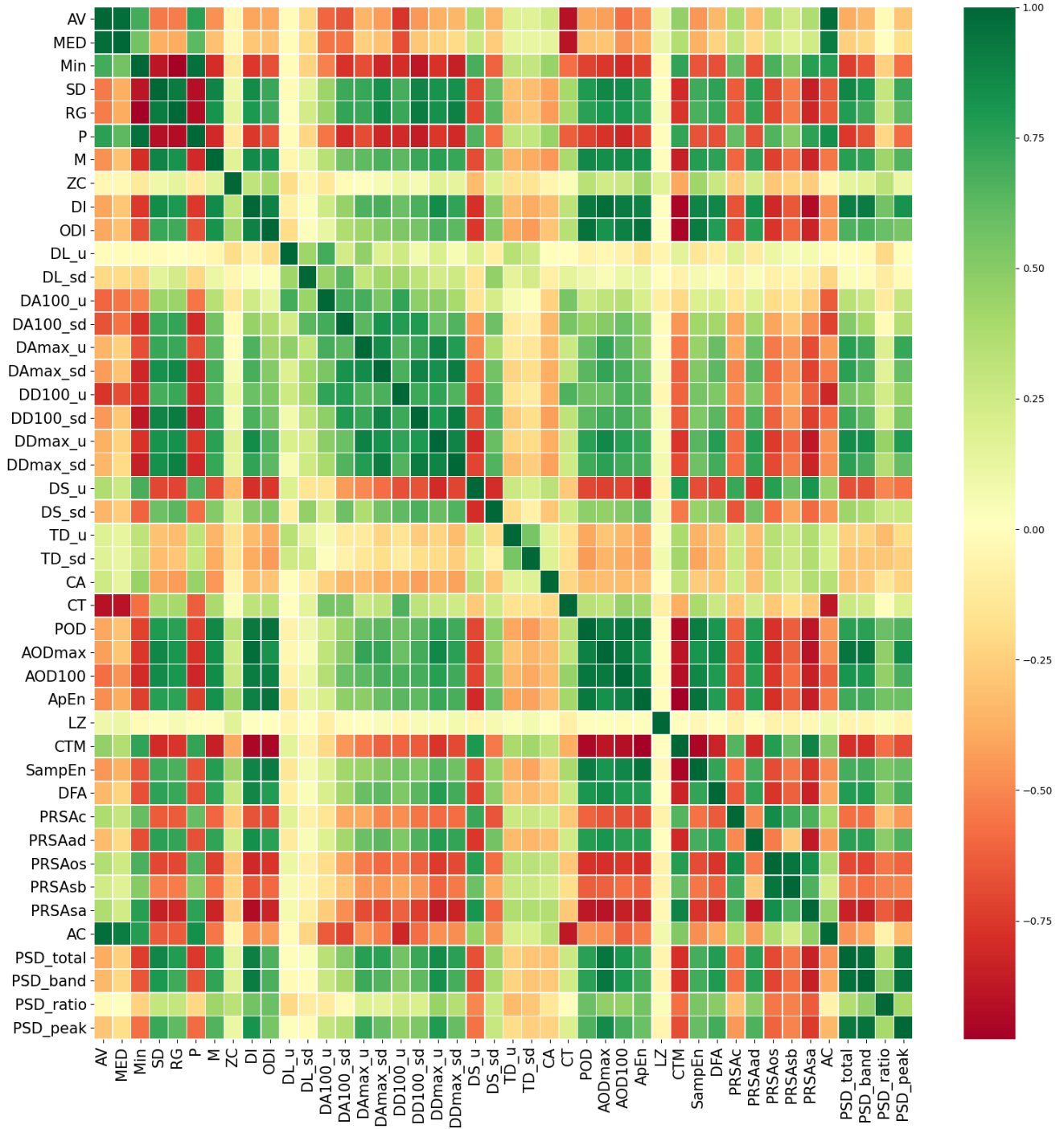

**Figure S3:** HeatMap of correlation of extracted features.

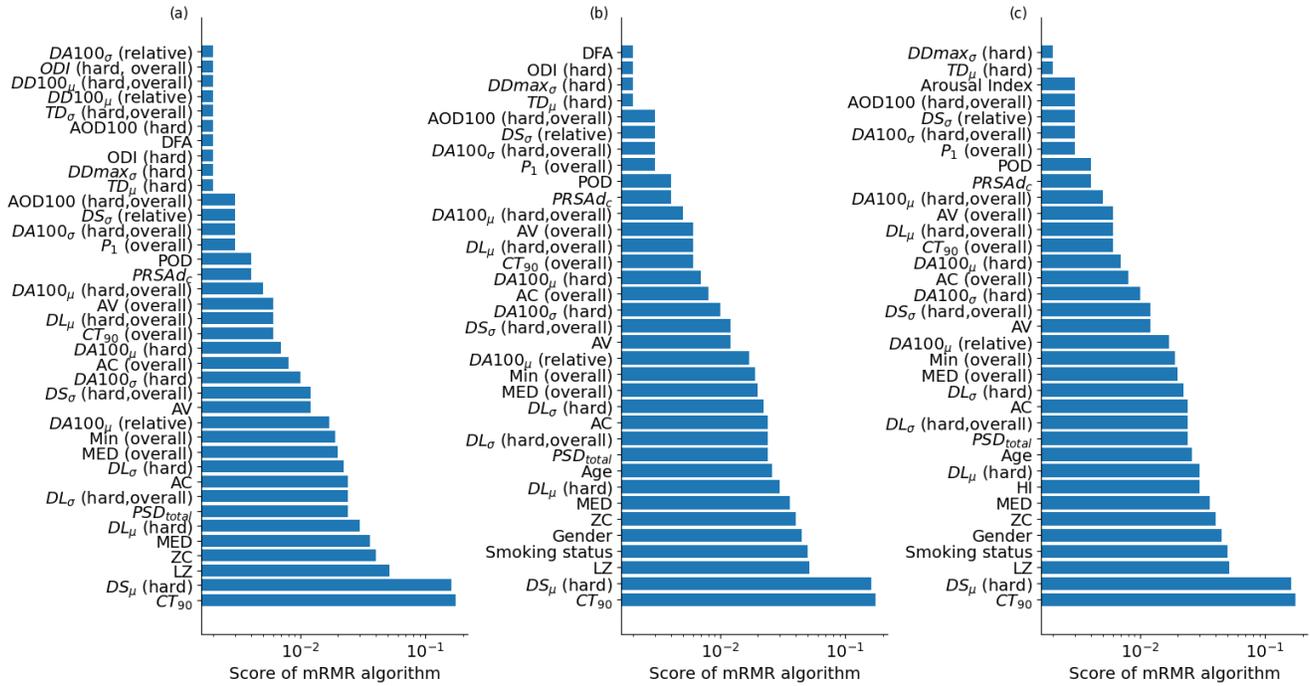

**Figure S4:** Scores of mRMR algorithm. The relevance of each feature, $I(x_i, c)$ is represented. Panels: (a) model 2; (b) model 3; (c) model 4.

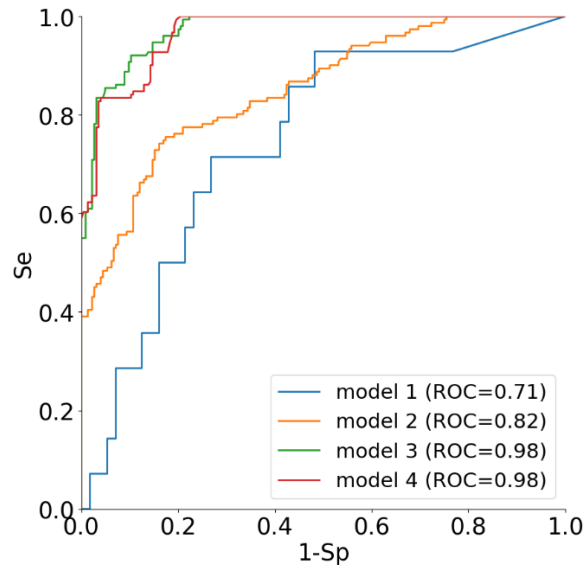

**Figure S5:** ROC curves for each of the 4 models, on the test set. Results are presented for the per window classification.

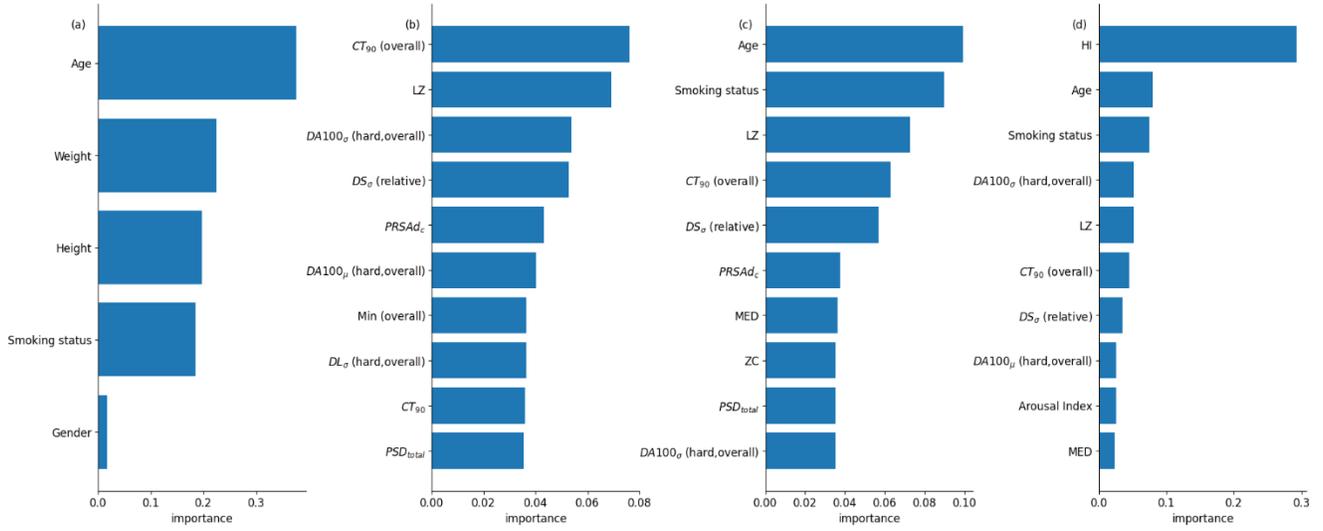

**Figure S6:** Feature importance determined by the RF classifier. The most important features are shown for (a) model 1; (b) model 2; (c) model 3 and (d) model 4.

|  |  | **Biomarker** | **Definition** | **Unit** |
|---|---|---|---|---|
| **General statistics** | 1 | AV | Blood oxygen saturation ($SpO_2$) mean | % |
|  | 2 | MED | $SpO_2$ median | % |
|  | 3 | Min | $SpO_2$ min | % |
|  | 4 | SD | $SpO_2$ standard deviation | % |
|  | 5 | RG | $SpO_2$ range | % |
|  | 6 | Px | $x^{th}$ percentile $SpO_2$ value. | % |
|  | 7 | Mx | Percentage of the signal at least x% below-median oxygen saturation. | % |
|  | 8 | ZCx | Number of zero-crossing points at the x% $SpO_2$ level. | nu |
|  | 9 | ΔIx | Delta index. | % |
| **Complexity** | 10 | ApEn | Approximate entropy. | nu |
|  | 11 | LZ | Lempel-Ziv complexity. | nu |
|  | 12 | $CTM_\rho$ | Central tendency measure with radius ρ. | nu |
|  | 13 | SampEn | Sample entropy. | nu |
|  | 14 | DFA | Detrented Fluctuation Analysis. | % |
| **Periodicity** | 15 | $PRSAd_c$ | Phase-Rectified Signal Averaging (PRSA) capacity, with d the fragment duration. | % |
|  | 16 | $PRSAd_{ad}$ | PRSA amplitude difference, with d the fragment duration. | % |
|  | 17 | $PRSAd_{os}$ | PRSA overall slope. With d the fragment duration. | %/sec |
|  | 18 | $PRSAd_{sb}$ | PRSA slope before the anchor point. With d the fragment duration. | %/sec |
|  | 19 | $PRSAd_{sa}$ | PRSA slope after the anchor point, with d the fragment duration. | %/sec |
|  | 20 | AC | Autocorrelation | $\%^2$ |
|  | 21 | PSD_total | The integral of the Power Spectral Density (PSD) function . | % |
|  | 22 | PSD_band | The integral of the PSD function within the band 0.014 − 0.033 Hz. | % |
|  | 23 | PSD_ratio | The integral of the PSD function within the band 0.014 − 0.033 Hz with respect to the total integral. | nu |
|  | 24 | PSD_peak | Peak amplitude of the PSD function within the band 0.014 − 0.033 Hz. | % |
| **Desaturation measures** | 25 | ODIx | The oxygen desaturation index. | event/h |
| | 26 | $DL_\mu$ | Mean of desaturations length. | sec |
| | 27 | $DL_\sigma$ | Standard deviation of desaturations length. | $sec^2$ |
| | 28 | $DDmax_\mu$ | Mean of desaturations depth. | % |
| | 29 | $DDmax_\sigma$ | Standard deviation of desaturations depth. | $\%^2$ |
| | 30 | $DD100_\mu$ | Mean of desaturations depth using 100% $SpO_2$ level as baseline. | % |
| | 31 | $DD100_\sigma$ | Standard deviation of desaturation depth using 100% $SpO_2$ level as baseline. | $\%^2$ |
| | 32 | $DS_\mu$ | Mean of the desaturation slope. | %/sec |
| | 33 | $DS_\sigma$ | Standard deviation of the desaturation slope. | $(\%/sec)^2$ |
| | 34 | $DAmax_\mu$ | Desaturation area defined as the mean of the desaturation areas using the maximum $SpO_2$ value in each desaturation event as the baseline. | %*sec |
| | 35 | $DAmax_\sigma$ | Standard deviation of desaturation area. | $(\%*sec)^2$ |
| | 36 | $DA100_\mu$ | Desaturation area: mean of desaturation area under the 100% $SpO_2$ level as baseline. | %*sec |
| | 37 | $DA100_\sigma$ | Standard deviation of desaturation area under the 100% $SpO_2$ level as baseline. | $(\%*sec)^2$ |
| | 38 | $TD_\mu$ | Mean of time between two consecutive desaturation events. | sec |
| | 39 | $TD_\sigma$ | Standard deviation of time between two consecutive desaturation events. | $sec^2$ |
| **Hypoxic burden** | 40 | PODx | Time of oxygen desaturation event, normalized by the total recording time. | sec |
| | 41 | AODmax | The area under the oxygen desaturation event curve, using the maximum $SpO_2$ value as baseline and normalized by the total recording time. | % |
| | 42 | AOD100 | Cumulative area of desaturations under the 100% $SpO_2$ level as baseline and normalized by the total recording time. | % |
| | 43 | CTx | Cumulative time below the x% oxygen saturation level. | % |
| | 44 | CAx | Integral of $SpO_2$ below the x $SpO_2$ level normalized by the total recording time. | % |
| **Demographi** | 45 | *Gender* | | n.u. |
| | 46 | *Age* | | years |
| | 47 | *Weight* | | kg |
| | 48 | *Height* | | cm |

|  | 49 | *Smoking status* | Whether smoker, ex-smoker, or non-smoker. | n.u. |
|---|---|---|---|---|
| **PSG** | 50 | *AHI* | Apnea-hypopnea index. | Events /h |
|  | 51 | *AI* | Apnea index. | Events /h |
|  | 52 | *HI* | Hypopnea index. | Events /h |
|  | 53 | *N1* | Percentage of time spent in stage NREM . | % of Total Sleep Time (TST) |
|  | 54 | *N2* | Percentage of time spent in stage NREM 2. | % of TST |
|  | 55 | *N3* | Percentage of time spent in stage NREM 3. | % of TST |
|  | 56 | *REM* | Percentage of time spent in stage REM. | % of TST |
|  | 57 | *Arousal* | Arousal index. | Arousal/h |
|  | 58 | *SE* | Sleep efficiency. | % of TST |

**Table S1**: List of biomarkers used, as defined in (Levy *et al* 2020).

|  |  | Name | non-COPD (m = 1088) | GOLD1 (m = 138) | GOLD2 (m = 306) | GOLD3 (m = 61) | GOLD4 (m = 6) |
|---|---|---|---|---|---|---|---|
| General statistics | 1 | AV | 93.24 ± 2.66 | 91.93 ± 3.42 | 90.78 ± 4.40 | 90.19 ± 4.26 | 90.55 ± 0.68 |
|  | 2 | MED | 93.00 ± 3.00 | 92.00 ± 3.00 | 91.00 ± 4.00 | 90.00 ± 3.00 | 91.00 ± 1.00 |
|  | 3 | Min | 85.00 ± 9.00 | 86.00 ± 8.00 | 83.00 ± 8.00 | 82.00 ± 8.75 | 81.00 ± 4.00 |
|  | 4 | SD | 1.66 ± 1.47 | 1.57 ± 1.09 | 1.78 ± 1.34 | 2.07 ± 1.31 | 3.22 ± 0.39 |
|  | 5 | RG | 12.00 ± 9.00 | 11.00 ± 9.00 | 12.00 ± 8.00 | 15.00 ± 8.75 | 16.00 ± 3.50 |
|  | 6 | Px | 88.00 ± 7.00 | 88.00 ± 7.00 | 85.00 ± 7.00 | 85.00 ± 9.00 | 83.00 ± 2.50 |
|  | 7 | Mx | 6.03 ± 14.54 | 5.39 ± 13.12 | 7.11 ± 14.54 | 9.30 ± 16.35 | 29.15 ± 4.62 |
|  | 8 | ZCx | 200.00 ± 96.00 | 180.00 ± 99.00 | 167.00 ± 121.00 | 156.00 ± 108.75 | 362.00 ± 52.50 |
|  | 9 | ΔIx | 0.86 ± 1.03 | 0.71 ± 0.86 | 0.80 ± 0.78 | 1.02 ± 0.65 | 2.93 ± 0.49 |
| Complexity | 10 | ApEn | 0.37 ± 0.26 | 0.33 ± 0.23 | 0.36 ± 0.22 | 0.41 ± 0.23 | 0.74 ± 0.04 |
|  | 11 | LZ | 122.00 ± 0.00 | 121.00 ± 1.00 | 122.00 ± 0.00 | 122.00 ± 0.00 | 122.00 ± 0.00 |
|  | 12 | $CTM_\rho$ | 0.81 ± 0.17 | 0.83 ± 0.15 | 0.81 ± 0.15 | 0.79 ± 0.16 | 0.47 ± 0.06 |
|  | 13 | SampEn | 0.16 ± 0.15 | 0.15 ± 0.17 | 0.17 ± 0.16 | 0.20 ± 0.16 | 0.45 ± 0.08 |
|  | 14 | DFA | 1.68 ± 2.12 | 1.44 ± 1.68 | 1.47 ± 1.23 | 2.05 ± 1.64 | 4.61 ± 1.33 |
| Periodicity | 15 | $PRSAd_c$ | -0.52 ± 0.07 | -0.51 ± 0.03 | -0.52 ± 0.04 | -0.52 ± 0.06 | -0.62 ± 0.03 |
|  | 16 | $PRSAd_{ad}$ | 2.14 ± 2.33 | 1.87 ± 2.28 | 1.96 ± 1.45 | 2.12 ± 1.18 | 4.72 ± 0.36 |
|  | 17 | $PRSAd_{os}$ | -0.14 ± 0.14 | -0.11 ± 0.11 | -0.13 ± 0.10 | -0.14 ± 0.08 | -0.31 ± 0.02 |
|  | 18 | $PRSAd_{sb}$ | -0.06 ± 0.14 | -0.04 ± 0.13 | -0.05 ± 0.11 | -0.07 ± 0.05 | -0.23 ± 0.02 |
|  | 19 | $PRSAd_{sa}$ | -0.06 ± 0.13 | -0.04 ± 0.11 | -0.04 ± 0.10 | -0.07 ± 0.04 | -0.21 ± 0.02 |
|  | 20 | AC | 103493.15 ± 6446.91 | 101128.01 ± 7902.74 | 97676.77 ± 9956.38 | 97704.64 ± 9373.68 | 97851.50 ± 1147.92 |
|  | 21 | PSD_total | 86.99 ± 11.38 | 84.23 ± 6.56 | 84.13 ± 9.19 | 85.55 ± 6.92 | 101.67 ± 6.54 |
|  | 22 | PSD_band | 3.80 ± 4.51 | 2.91 ± 3.56 | 3.18 ± 3.13 | 4.25 ± 2.82 | 10.79 ± 3.12 |
|  | 23 | PSD_ratio | 0.04 ± 0.04 | 0.03 ± 0.04 | 0.04 ± 0.03 | 0.05 ± 0.03 | 0.11 ± 0.02 |
|  | 24 | PSD_peak | 0.08 ± 0.10 | 0.06 ± 0.09 | 0.06 ± 0.07 | 0.09 ± 0.06 | 0.21 ± 0.07 |
| Desaturations relative | 25 | ODIx | 14.21 ± 32.68 | 7.11 ± 28.89 | 9.95 ± 22.87 | 17.76 ± 31.62 | 84.79 ± 15.19 |
|  | 26 | $DL_\mu$ | 39.11 ± 17.07 | 43.50 ± 22.36 | 39.17 ± 18.49 | 37.51 ± 10.22 | 26.73 ± 0.74 |
|  | 27 | $DL_\sigma$ | 19.32 ± 13.46 | 20.24 ± 11.10 | 20.84 ± 13.98 | 22.16 ± 7.35 | 11.45 ± 1.51 |
|  | 28 | $DDmax_\mu$ | 336.83 ± 180.71 | 383.89 ± 323.28 | 431.60 ± 243.09 | 431.38 ± 227.27 | 307.54 ± 26.93 |

| | | | | | | |
|---|---|---|---|---|---|---|
| | 29 | DDmax$_\sigma$ | 161.38 ± 146.14 | 195.59 ± 133.07 | 218.34 ± 245.14 | 225.75 ± 184.08 | 146.09 ± 13.50 |
| | 30 | DD100$_\mu$ | 112.33 ± 88.32 | 109.29 ± 99.12 | 101.95 ± 74.13 | 95.17 ± 78.31 | 111.01 ± 13.02 |
| | 31 | DD100$_\sigma$ | 79.05 ± 92.77 | 92.84 ± 97.60 | 79.82 ± 95.17 | 79.60 ± 78.28 | 76.09 ± 14.21 |
| | 32 | DS$_\mu$ | 10.28 ± 3.72 | 11.00 ± 4.32 | 12.52 ± 5.11 | 11.57 ± 5.60 | 13.09 ± 0.49 |
| | 33 | DS$_\sigma$ | 1.68 ± 1.80 | 1.54 ± 1.65 | 1.72 ± 1.53 | 1.58 ± 1.35 | 2.44 ± 0.35 |
| | 34 | DAmax$_\mu$ | 4.50 ± 2.59 | 4.03 ± 1.71 | 4.10 ± 1.89 | 4.14 ± 1.24 | 6.21 ± 0.44 |
| | 35 | DAmax$_\sigma$ | 1.50 ± 2.05 | 1.15 ± 1.46 | 1.27 ± 1.71 | 1.44 ± 1.04 | 2.35 ± 0.27 |
| | 36 | DA100$_\mu$ | -0.19 ± 0.13 | -0.17 ± 0.08 | -0.18 ± 0.10 | -0.17 ± 0.10 | -0.31 ± 0.02 |
| | 37 | DA100$_\sigma$ | 0.09 ± 0.07 | 0.07 ± 0.03 | 0.09 ± 0.05 | 0.09 ± 0.04 | 0.08 ± 0.02 |
| | 38 | TD$_\mu$ | 147.59 ± 20.6 | 242.08 ± 51.63 | 184.69 ± 71.23 | 145.38 ± 225.99 | 42.35 ± 6.46 |
| | 39 | TD$_\sigma$ | 215.92 ± 22.7 | 271.85 ± 102.32 | 230.33 ± 75.23 | 168.40 ± 292.96 | 16.16 ± 40.14 |
| Desaturation hard | 40 | ODIx | 7.11 ± 27.00 | 12.79 ± 30.47 | 16.11 ± 30.32 | 21.84 ± 34.93 | 69.63 ± 15.42 |
| | 41 | DL$_\mu$ | 13.96 ± 15.30 | 19.14 ± 22.79 | 27.48 ± 58.53 | 25.15 ± 30.12 | 17.54 ± 2.88 |
| | 42 | DL$_\sigma$ | 6.17 ± 14.17 | 10.02 ± 28.89 | 30.47 ± 155.28 | 16.50 ± 67.94 | 8.43 ± 3.60 |
| | 43 | DDmax$_\mu$ | 177.03 ± 217.41 | 236.48 ± 327.57 | 374.80 ± 805.72 | 325.03 ± 363.62 | 242.85 ± 39.27 |
| | 44 | DDmax$_\sigma$ | 86.25 ± 219.49 | 140.84 ± 543.94 | 464.51 ± 1975.71 | 221.07 ± 923.77 | 142.30 ± 46.19 |
| | 45 | DD100$_\mu$ | 31.10 ± 69.25 | 29.19 ± 85.07 | 75.02 ± 221.83 | 57.13 ± 117.47 | 65.19 ± 14.54 |
| | 46 | DD100$_\sigma$ | 27.36 ± 80.35 | 30.96 ± 120.19 | 122.97 ± 596.85 | 65.01 ± 266.61 | 64.47 ± 26.37 |
| | 47 | DS$_\mu$ | 11.95 ± 2.37 | -0.05 ± 0.02 | -0.03 ± 0.04 | -0.26 ± 0.07 | -0.30 ± 0.01 |
| | 48 | DS$_\sigma$ | 1.04 ± 2.20 | 0.03 ± 0.06 | 0.00 ± 0.05 | 0.04 ± 0.03 | 0.06 ± 0.02 |
| | 49 | DAmax$_\mu$ | 2.12 ± 2.80 | 1.75 ± 2.69 | 2.35 ± 2.30 | 1.95 ± 2.27 | 4.04 ± 0.37 |
| | 50 | DAmax$_\sigma$ | 1.24 ± 2.46 | 0.99 ± 1.83 | 1.60 ± 1.74 | 1.23 ± 1.26 | 2.28 ± 0.46 |
| | 51 | DA100$_\mu$ | 0.00 ± 0.1 | 2.64 ± 1.32 | 3.45 ± 1.21 | 3.29 ± 2.54 | 5.23 ± 1.23 |
| | 52 | DA100$_\sigma$ | 0.00 ± 0.2 | 0.25 ± 0.99 | 0.65 ± 1.2 | 0.85 ± 0.80 | 1.00 ± 0.65 |
| | 53 | TD$_\mu$ | 89.36 ± 10.2 | 103.65 ± 13.52 | 113.29 ± 8.65 | 98.97 ± 7.23 | 51.69 ± 11.52 |
| | 54 | TD$_\sigma$ | 89.94 ± 12.6 | 98.25 ± 21.32 | 144.56 ± 22.25 | 102.37 ± 18.96 | 25.77 ± 38.00 |
| Hypoxic | 55 | PODx | 0.16 ± 0.33 | 0.12 ± 0.27 | 0.12 ± 0.24 | 0.19 ± 0.30 | 0.63 ± 0.13 |
| | 56 | AODmax | 0.43 ± 1.33 | 0.28 ± 0.94 | 0.30 ± 0.85 | 0.46 ± 0.92 | 2.62 ± 0.76 |

| | | Name | non-COPD ($n = 280$) | GOLD1 ($n = 14$) | GOLD2 ($n = 46$) | GOLD3 ($n = 9$) | GOLD4 ($n = 1$) |
|---|---|---|---|---|---|---|---|
| | 57 | AOD100 | 1.34 ± 3.26 | 1.09 ± 3.31 | 1.41 ± 2.98 | 1.82 ± 3.05 | 6.99 ± 1.67 |
| | 58 | CTx | 5.95 ± 24.60 | 22.24 ± 53.80 | 39.42 ± 81.60 | 53.28 ± 57.55 | 38.80 ± 11.85 |
| | 59 | CAx | 0.62 ± 0.53 | 0.60 ± 0.34 | 0.67 ± 0.53 | 0.75 ± 0.54 | 1.34 ± 0.21 |

**Table S2:** Median (MED) and interquartile range (±IQR) descriptive statistics of the population sample studied for all the oximetry biomarkers, per window. The number of patients is represented by $m$.

| | | Name | non-COPD ($n = 280$) | GOLD1 ($n = 14$) | GOLD2 ($n = 46$) | GOLD3 ($n = 9$) | GOLD4 ($n = 1$) |
|---|---|---|---|---|---|---|---|
| General statistics | 1 | AV | 93.22 ± 2.52 | 92.04 ± 3.16 | 91.04 ± 3.90 | 89.90 ± 3.33 | 90.73 ± 0.00 |
| | 2 | MED | 93.00 ± 2.25 | 92.00 ± 3.00 | 91.00 ± 4.00 | 90.00 ± 2.00 | 91.00 ± 0.00 |
| | 3 | Min | 82.00 ± 11.00 | 84.00 ± 12.00 | 81.00 ± 11.00 | 77.00 ± 13.00 | 78.00 ± 0.00 |
| | 4 | SD | 1.87 ± 1.48 | 1.74 ± 0.93 | 1.96 ± 1.29 | 2.19 ± 1.87 | 3.06 ± 0.00 |
| | 5 | RG | 16.00 ± 11.00 | 14.00 ± 8.00 | 16.00 ± 11.00 | 19.00 ± 10.00 | 19.00 ± 0.00 |
| | 6 | Px | 87.00 ± 8.00 | 87.00 ± 8.00 | 85.00 ± 8.00 | 84.00 ± 14.00 | 83.00 ± 0.00 |
| | 7 | Mx | 8.47 ± 13.27 | 6.31 ± 7.03 | 7.85 ± 13.05 | 7.04 ± 14.11 | 22.12 ± 0.00 |
| | 8 | ZCx | 648.50 ± 293.25 | 676.00 ± 269.00 | 574.00 ± 368.00 | 639.00 ± 298.00 | 1244.00 ± 0.00 |
| | 9 | ΔIx | 0.95 ± 0.93 | 0.70 ± 0.93 | 0.83 ± 0.62 | 1.00 ± 0.40 | 2.64 ± 0.00 |
| Complexity | 10 | ApEn | 0.39 ± 0.21 | 0.37 ± 0.31 | 0.45 ± 0.45 | 0.69 ± 0.20 | 0.84 ± 0.00 |
| | 11 | LZ | 229.00 ± 7.00 | 229.50 ± 6.75 | 230.00 ± 6.75 | 229.00 ± 5.00 | 233.00 ± 0.00 |
| | 12 | $CTM_\rho$ | 0.80 ± 0.14 | 0.83 ± 0.14 | 0.81 ± 0.12 | 0.78 ± 0.10 | 0.51 ± 0.00 |
| | 13 | SampEn | 0.25 ± 0.19 | 0.13 ± 0.09 | 0.15 ± 0.12 | 0.19 ± 0.14 | 0.45 ± 0.00 |
| | 14 | DFA | 1.45 ± 1.91 | 1.33 ± 1.75 | 0.98 ± 0.97 | 1.56 ± 0.94 | 3.77 ± 0.00 |
| Periodicity | 15 | $PRSAd_c$ | -0.52 ± 0.08 | -0.51 ± 0.03 | -0.52 ± 0.04 | -0.52 ± 0.04 | -0.62 ± 0.00 |
| | 16 | $PRSAd_{ad}$ | 2.34 ± 2.29 | 1.73 ± 2.39 | 1.96 ± 1.52 | 2.26 ± 0.86 | 4.56 ± 0.00 |
| | 17 | $PRSAd_{os}$ | -0.15 ± 0.14 | -0.12 ± 0.10 | -0.13 ± 0.11 | -0.15 ± 0.06 | -0.30 ± 0.00 |
| | 18 | $PRSAd_{sb}$ | -0.08 ± 0.14 | -0.04 ± 0.12 | -0.06 ± 0.12 | -0.07 ± 0.03 | -0.22 ± 0.00 |
| | 19 | $PRSAd_{sa}$ | -0.07 ± 0.13 | -0.05 ± 0.11 | -0.05 ± 0.10 | -0.08 ± 0.03 | -0.20 ± 0.00 |
| | 20 | AC | 102991.64 ± 5755.47 | 101600.64 ± 7084.13 | 97932.80 ± 9283.94 | 97345.13 ± 8093.00 | 97935.83 ± 0.00 |
| | 21 | PSD_total | 103.17 ± 23.95 | 98.30 ± 13.68 | 99.10 ± 18.26 | 100.78 ± 12.76 | 129.57 ± 0.00 |
| | 22 | PSD_band | 8.04 ± 9.18 | 6.20 ± 6.75 | 6.92 ± 5.59 | 8.20 ± 2.97 | 19.89 ± 0.00 |

|  | | | | | | |
|---|---|---|---|---|---|---|
|  | 23 | PSD_ratio | 0.08 ± 0.07 | 0.06 ± 0.06 | 0.07 ± 0.05 | 0.08 ± 0.02 | 0.15 ± 0.00 |
|  | 24 | PSD_peak | 0.05 ± 0.07 | 0.05 ± 0.06 | 0.04 ± 0.04 | 0.05 ± 0.02 | 0.14 ± 0.00 |
| **Desaturations relative** | 25 | ODIx | 16.01 ± 28.62 | 8.59 ± 29.89 | 10.88 ± 17.71 | 19.37 ± 17.29 | 73.57 ± 0.00 |
|  | 26 | $DL_\mu$ | 39.19 ± 14.89 | 45.05 ± 20.65 | 40.05 ± 14.30 | 37.33 ± 9.54 | 26.41 ± 0.00 |
|  | 27 | $DL_\sigma$ | 21.75 ± 9.72 | 22.02 ± 4.14 | 23.27 ± 9.31 | 23.46 ± 2.56 | 12.02 ± 0.00 |
|  | 28 | $DDmax_\mu$ | 4.82 ± 2.45 | 4.24 ± 1.27 | 4.16 ± 1.47 | 4.38 ± 1.01 | 6.00 ± 0.00 |
|  | 29 | $DDmax_\sigma$ | 1.95 ± 2.01 | 1.64 ± 1.16 | 1.55 ± 1.51 | 1.55 ± 0.86 | 2.38 ± 0.00 |
|  | 30 | $DD100_\mu$ | 10.65 ± 3.41 | 11.14 ± 4.33 | 12.68 ± 4.35 | 12.60 ± 5.56 | 13.05 ± 0.00 |
|  | 31 | $DD100_\sigma$ | 2.15 ± 1.78 | 2.03 ± 1.66 | 2.06 ± 1.67 | 2.02 ± 1.63 | 2.46 ± 0.00 |
|  | 32 | $DS_\mu$ | -0.20 ± 0.11 | -0.17 ± 0.07 | -0.18 ± 0.08 | -0.17 ± 0.07 | -0.32 ± 0.00 |
|  | 33 | $DS_\sigma$ | 0.10 ± 0.06 | 0.08 ± 0.03 | 0.10 ± 0.04 | 0.09 ± 0.03 | 0.10 ± 0.00 |
|  | 34 | $DAmax_\mu$ | 116.72 ± 70.59 | 130.58 ± 79.99 | 104.28 ± 48.88 | 100.70 ± 90.04 | 107.18 ± 0.00 |
|  | 35 | $DAmax_\sigma$ | 95.91 ± 84.81 | 107.60 ± 91.24 | 94.74 ± 67.20 | 82.63 ± 68.66 | 76.23 ± 0.00 |
|  | 36 | $DA100_\mu$ | 349.32 ± 161.19 | 365.79 ± 320.92 | 473.03 ± 232.18 | 476.92 ± 215.71 | 298.25 ± 0.00 |
|  | 37 | $DA100_\sigma$ | 180.29 ± 130.25 | 211.52 ± 113.69 | 288.22 ± 219.02 | 216.29 ± 163.96 | 145.31 ± 0.00 |
|  | 38 | $TD_\mu$ | 202.42 ± 253.32 | 379.89 ± 529.33 | 301.85 ± 759.74 | 174.67 ± 190.91 | 48.32 ± 0.00 |
|  | 39 | $TD_\sigma$ | 438.59 ± 456.39 | 709.48 ± 752.42 | 549.02 ± 1329.85 | 300.00 ± 356.19 | 59.82 ± 0.00 |
| **Desaturation hard** | 40 | ODIx | 42.26 ± 21.54 | 35.73 ± 25.31 | 33.84 ± 26.37 | 42.74 ± 18.75 | 81.57 ± 0.00 |
|  | 41 | $DL_\mu$ | 28.37 ± 12.83 | 33.74 ± 17.91 | 29.98 ± 24.22 | 29.83 ± 3.96 | 17.21 ± 0.00 |
|  | 42 | $DL_\sigma$ | 43.03 ± 48.38 | 63.84 ± 58.27 | 72.55 ± 90.45 | 80.19 ± 97.69 | 9.53 ± 0.00 |
|  | 43 | $DDmax_\mu$ | 2.35 ± 1.91 | 1.71 ± 1.19 | 2.05 ± 1.16 | 2.05 ± 0.69 | 4.01 ± 0.00 |
|  | 44 | $DDmax_\sigma$ | 1.80 ± 1.88 | 1.21 ± 1.29 | 1.52 ± 1.40 | 1.69 ± 0.72 | 2.55 ± 0.00 |
|  | 45 | $DD100_\mu$ | 9.29 ± 3.29 | 10.30 ± 3.28 | 11.74 ± 4.52 | 12.00 ± 4.65 | 12.84 ± 0.00 |
|  | 46 | $DD100_\sigma$ | 1.73 ± 1.76 | 1.11 ± 1.29 | 1.50 ± 1.30 | 1.58 ± 0.84 | 2.38 ± 0.00 |
|  | 47 | $DS_\mu$ | -0.06 ± 0.03 | -0.05 ± 0.01 | -0.03 ± 0.10 | -0.28 ± 0.16 | -0.35 ± 0.00 |
|  | 48 | $DS_\sigma$ | 0.00 ± 0.01 | 0.04 ± 0.08 | 0.00 ± 0.12 | 0.04 ± 0.05 | 0.06 ± 0.00 |
|  | 49 | $DAmax_\mu$ | 61.71 ± 54.42 | 74.23 ± 51.88 | 72.63 ± 73.81 | 80.40 ± 89.32 | 62.74 ± 0.00 |
|  | 50 | $DAmax_\sigma$ | 119.79 ± 135.73 | 172.79 ± 185.59 | 198.48 ± 345.20 | 319.33 ± 252.13 | 62.36 ± 0.00 |

|  | | Name | non-COPD ($n = 280$) | GOLD1 ($n = 14$) | GOLD2 ($n = 46$) | GOLD3 ($n = 9$) | GOLD4 ($n = 1$) |
|---|---|---|---|---|---|---|---|
|  | 51 | $DA100_\mu$ | 267.34 ± 157.12 | 348.86 ± 127.91 | 398.72 ± 367.12 | 422.01 ± 294.94 | 222.34 ± 0.00 |
|  | 52 | $DA100_\sigma$ | 395.53 ± 475.19 | 604.64 ± 493.51 | 828.10 ± 1200.12 | 978.25 ± 848.20 | 137.96 ± 0.00 |
|  | 53 | $TD_\mu$ | 83.37 ± 43.50 | 96.68 ± 66.35 | 101.58 ± 63.37 | 81.39 ± 37.97 | 43.45 ± 0.00 |
|  | 54 | $TD_\sigma$ | 122.72 ± 114.36 | 163.59 ± 159.29 | 167.63 ± 146.61 | 163.29 ± 134.33 | 30.94 ± 0.00 |
| Hypoxic | 55 | PODx | 0.37 ± 0.16 | 0.35 ± 0.12 | 0.42 ± 0.13 | 0.40 ± 0.07 | 0.39 ± 0.00 |
| Hypoxic | 56 | AODmax | 0.76 ± 0.74 | 0.62 ± 0.29 | 0.85 ± 0.64 | 0.82 ± 0.99 | 1.42 ± 0.00 |
| Hypoxic | 57 | AOD100 | 3.37 ± 2.12 | 3.68 ± 1.45 | 4.97 ± 3.15 | 4.53 ± 3.55 | 5.04 ± 0.00 |
| Hypoxic | 58 | CTx | 8.22 ± 22.21 | 21.43 ± 51.01 | 40.69 ± 71.39 | 62.66 ± 35.67 | 38.96 ± 0.00 |
| Hypoxic | 59 | CAx | 0.68 ± 0.49 | 0.63 ± 0.25 | 0.76 ± 0.46 | 0.83 ± 0.58 | 1.23 ± 0.00 |

**Table S3:** Median (MED) and interquartile range (±IQR) descriptive statistics of the population sample studied for all the oximetry biomarkers computed on the overall signal. The number of windows is represented by $n$.

|  |  | Name | non-COPD ($n = 280$) | GOLD1 ($n = 14$) | GOLD2 ($n = 46$) | GOLD3 ($n = 9$) | GOLD4 ($n = 1$) |
|---|---|---|---|---|---|---|---|
| Demographics | 1 | $Gender$ | Male : 206 (74%) Female: 74 (26%) | Male : 11 (79%) Female: 3 (21%) | Male : 41 (89%) Female: 5 (11%) | Male : 8 (89%) Female: 1 (11%) | Male : 1 (100%) Female: 0 (0%) |
| Demographics | 2 | $Age$ | 54.00 ± 19.00 | 61.00 ± 13.00 | 65.00 ± 11.00 | 65.00 ± 9.00 | 78.00 ± 0.00 |
| Demographics | 3 | $Weight$ | 82.00 ± 20.00 | 86.00 ± 31.00 | 84.00 ± 16.00 | 81.00 ± 20.00 | 83.00 ± 0.00 |
| Demographics | 4 | $Height$ | 170.00 ± 13.00 | 170.00 ± 18.00 | 168.00 ± 10.00 | 166.00 ± 9.00 | 169.00 ± 0.00 |
| Demographics | 5 | $Smoking\ status$ | No smoker: 83 (24%) Smoker: 139 (40%) Ex-smoker: 58 (16%) | No smoker: 0 (0%) Smoker: 9 (64%) Ex-smoker: 5 (36%) | No smoker: 0 (0%) Smoker: 29 (63%) Ex-smoker: 17 (0%) | No smoker: 0 (0%) Smoker: 6 (66%) Ex-smoker: 3 (33%) | No smoker: 0 (0%) Smoker: 1 (0%) Ex-smoker: 0 (100%) |

**Table S4:** Median (MED) and interquartile range (±IQR) descriptive statistics of the population sample studied for all the demographic features. The number of patients is represented by $n$.

|  |  | Name | non-COPD ($n = 280$) | GOLD1 ($n = 14$) | GOLD2 ($n = 46$) | GOLD3 ($n = 9$) | GOLD4 ($n = 1$) |
|---|---|---|---|---|---|---|---|
| PSG | 1 | $AHI$ | 35.15 ± 42.67 | 31.00 ± 43.50 | 33.20 ± 45.80 | 57.90 ± 42.30 | 98.70 ± 0.00 |
|  | 2 | $AI$ | 11.25 ± 29.47 | 5.60 ± 37.60 | 8.70 ± 21.00 | 15.20 ± 36.50 | 86.40 ± 0.00 |
|  | 3 | $HI$ | 15.20 ± 0.00 | 10.80 ± 18.30 | 16.90 ± 12.40 | 16.90 ± 15.80 | 12.30 ± 0.00 |
|  | 4 | $N1$ | 13.60 ± 11.35 | 13.10 ± 11.50 | 16.70 ± 11.30 | 25.70 ± 23.20 | 37.00 ± 0.00 |
|  | 5 | $N2$ | 32.35 ± 13.62 | 28.90 ± 17.20 | 27.00 ± 11.80 | 25.90 ± 14.80 | 45.90 ± 0.00 |
|  | 6 | $N3$ | 35.45 ± 18.30 | 42.50 ± 12.50 | 38.90 ± 18.90 | 36.00 ± 32.40 | 11.00 ± 0.00 |
|  | 7 | $REM$ | 14.60 ± 9.28 | 13.90 ± 6.80 | 15.60 ± 9.50 | 10.70 ± 8.80 | 6.10 ± 0.00 |
|  | 8 | $Arousal$ | 29.50 ± 22.77 | 27.50 ± 12.60 | 30.70 ± 14.10 | 34.20 ± 23.20 | 45.80 ± 0.00 |
|  | 9 | $SE$ | 84.60 ± 16.53 | 81.70 ± 19.70 | 77.60 ± 15.80 | 67.70 ± 19.50 | 87.90 ± 0.00 |

**Table S5:** Median (MED) and interquartile range (±IQR) descriptive statistics of the population sample studied for all the PSG features.

|  | $Predictive\ non-COPD$ | $Predictive\ COPD$ |
|---|---|---|
| $Actual\ Non-COPD$ | 1046 | 72 |
| $Actual\ COPD$ | 42 | 469 |

**Table S6:** Confusion matrix for model 5, using RF classifier prediction on the test sets. Classification for individual windows is presented.

|  | RF | | | | LR | |
|---|---|---|---|---|---|---|
|  | Number estimators | Max depth | Minimum samples for split | Minimum sample for leaf | Learning rate ($*10^{-6}$) | Regularisation ($*10^{-2}$) |
| Model 1 | 50 ± 24 | 20 ± 10 | 4 ± 3.2 | 6 ± 1.2 | 4.6 ± 0.04 | 1.4 ± 1.0 |
| Model 2 | 120 ± 30 | 35 ± 12 | 12 ± 4 | 5 ± 1 | 1.3 ± 0.05 | 0.4 ± 0.1 |
| Model 3 | 110 ± 40 | 35 ± 8 | 13 ± 6 | 8 ± 3 | 0.3 ± 0.02 | 2.5 ± 0.9 |
| Model 4 | 140 ± 21 | 43 ± 15 | 18 ± 4 | 9 ± 4 | 1.5 ± 0.01 | 5.0 ± 0.9 |

**Table S7:** Mean and standard deviation of hyperparameters for RF and LR models, on the nested cross-fold validation.

|         | AUROC         | $F_1$         | Se            | Sp            | NPV           | PPV           |
|---------|---------------|---------------|---------------|---------------|---------------|---------------|
| Model 1 | 0.71 ± 0.13   | 0.61 ± 0.18   | 0.33 ± 0.24   | 0.89 ± 0.07   | 0.82 ± 0.05   | 0.35 ± 0.43   |
| Model 2 | 0.82 ± 0.03   | 0.75 ± 0.10   | 0.65 ± 0.11   | 0.82 ± 0.02   | 0.76 ± 0.06   | 0.72 ± 0.09   |
| Model 3 | <u>0.98 ± 0.01</u> | <u>0.90 ± 0.06</u> | 0.92 ± 0.03 | <u>0.90 ± 0.03</u> | <u>0.95 ± 0.02</u> | 0.84 ± 0.05 |
| Model 4 | <u>0.98 ± 0.01</u> | <u>0.90 ± 0.05</u> | <u>0.93 ± 0.04</u> | <u>0.90 ± 0.04</u> | <u>0.95 ± 0.01</u> | <u>0.85 ± 0.04</u> |

**Table S8:** Results of the nested cross-fold (outer-loop, test sets) for the classifiers with their hyperparameters optimized during cross-fold validation (inner-loop). The median and standard deviation of each performance measure over the five outer loops is presented. Results are presented for the per window classification.

**Supplementary Note 1:**

The GOLD was graded using post-bronchodilator % of predicted $FEV_1$ values: GOLD 1: $FEV1 \geq 80\%$; GOLD 2: $79\% \geq FEV1 \geq 50\%$; GOLD 3: $49\% \geq FEV1 \geq 30\%$; GOLD 4: $29\% \geq FEV1$ (Vogelmeier *et al* 2017).

**Supplementary Note 2:**

For Random Forests classifier, the grid focused on:

- Number of estimators (100, 110, 120, 150, 200, 250, 300)
- Number of features to consider at every split (could be all features or just the square of overall features)
- Maximum number of levels in the tree (from 10 to 110, with a pace of 10)
- Minimum number of samples required to split a node (2,5,10)
- Minimum number of samples required at each leaf node (1,2,4)
- Enable/Disable bootstrap

The parameters were tested for all possible combinations.